\documentclass[12pt]{article}

\usepackage[english]{babel}
\usepackage[utf8]{inputenc}
\usepackage{amsthm,amsmath}

\usepackage[top=1in, bottom=1in, left=1in, right=1in]{geometry}
\usepackage{amsmath}
\usepackage{mathtools}
\usepackage{booktabs}
\usepackage{amssymb}
\usepackage{array}
\usepackage{graphicx}
\usepackage{listings}
\usepackage{tikz}
\usepackage{enumerate}
\usepackage{enumitem}
\usepackage{bbm}
\usepackage{multirow}
\usepackage[bottom]{footmisc}
\usepackage{geometry}

\usepackage{setspace}

\usepackage[T1]{fontenc}
\usepackage{lipsum}
\usepackage{multicol}
\usepackage{changepage}

\usepackage[natbibapa]{apacite}
\usepackage{caption}
\usepackage{subfigure}

\usepackage{color-edits}
\addauthor{bx}{purple}  
\addauthor{td}{teal}
\addauthor{uv}{red}

\usepackage[11pt]{moresize}

\usepackage[font=footnotesize,labelfont=bf]{caption}

\usepackage[hidelinks]{hyperref} 
\hypersetup{ % remove the box around the link and make the link blue instead 
	colorlinks,
	linkcolor={blue},
	citecolor={blue},
	urlcolor={blue}
}
\usepackage[capitalize]{cleveref}

\usepackage[]{titlesec}

\usepackage{amsmath} % provides numberwithin (and lots more)
\usepackage{lipsum}

\newtheorem{theorem}{Theorem}
\newtheorem{lemma}{Lemma}
\newtheorem{proposition}{Proposition}

\newtheorem{observation}{Observation}

\begin{document}

\title{\Large Robust Contracting for Sequential Search\thanks{We thank the editor, four anonymous referees, Nageeb Ali, Ian Ball, Dan Barron, Dan Bernhardt, Matteo Camboni, George Georgiadis, Alexis Ghersengorin, Yingni Guo, Mallesh Pai, Alessandro Pavan, Mike Powell, Debraj Ray, Luis Rayo, P\"{e}llumb Reshidi, Ludvig Sinander, Bruno Strulovici, Jo\~{a}o Thereze, Nicholas Yannelis, Jidong Zhou, and various audiences at 2024 Stony Brook ICGT, 2024 SEA, 2025 EAYE, 2025 SAET, 2025 EC, 2026 AEA, 2026 CEME/NBER Decentralization Conference, CUFE, Duke, Northwestern, PSU, Queen's, Stanford, TAMU, Toronto, U Iowa, and UNC.}
}

\author{Th\'{e}o Durandard\thanks{University of Illinois, Urbana-Champaign. Email: theod@illinois.edu} \qquad Udayan Vaidya\thanks{Duke University. Email: udayan.vaidya@duke.edu} \qquad Boli Xu\thanks{University of Iowa. Email: boli-xu@uiowa.edu}}

\date{\today}

\maketitle

\begin{center}
\vspace{-2em}
\textcolor{blue}{\href{https://arxiv.org/abs/2504.17948}{Click Here For Latest Version}}\\
\vspace{1.6cm}
%\textbf{Abstract}
\end{center}

\begin{spacing}{1.5}

%\vspace{-2em}
\begin{abstract}
    A principal contracts with an agent who sequentially searches over projects to generate a prize. The principal is unaware of the agent's full set of projects and evaluates a contract by its worst-case performance. We characterize the principal's robustly optimal contracts, which are all debt-like: the agent is paid only when the prize exceeds a threshold. Debt is optimal because it preserves the agent's incentive to continue exploring, discouraging the agent from settling down for cheap, safe alternatives. Our predictions map to contracts used to incentivize innovation in practice.
\end{abstract}


\end{spacing}

\newpage

%----------------------------------------%
%----------------------------------------%
\section{Introduction}
\label{sec:intro}

Innovation requires exploring ideas whose value is only revealed after investigation. Because exploration is costly, a fundamental question is which ideas to pursue. Innovators must decide among prospects that differ in terms of their costs, potential upsides, and likelihoods of success. 
In many settings, however, the party conducting this exploration is not the one financing it.
The principal funding the innovation must instead rely on an expert agent to carry it out.
This delegated search introduces two frictions: the agent's search process may be difficult to monitor, and the agent may have superior information about the available alternatives. 
Faced with limited information and limited control, how can the principal design a contract that guides the agent to explore worthwhile ideas?

To study this question, we consider a contracting model with two key features: an agent who sequentially searches over projects, and a principal who is unaware of the agent's full set of alternatives. Each project is modeled as a ``Pandora's Box'' \citep{weitzman_optimal_1979}, characterized by a fixed inspection cost and a distribution of prizes. When offered a wage contract, the agent chooses the order in which to inspect projects, when to stop searching, and which prize to return to the principal. The agent is protected by limited liability, and his search process is unobservable. 

We assume the principal knows only one of the agent's available projects at the time of contracting, but she entertains the possibility of other projects being available. This known project could represent an author's sample chapters, an inventor's prototype, or an acquired startup's current business model that can be verified and vetted. The unknown projects reflect potential private information of the agent or simply alternatives that were unknown to both parties at the time of contracting, but may later become available. Given the agent's expertise and the uniqueness of each innovation, the principal may have little guidance on how to formulate a prior belief over this uncertainty. Following \citet{carroll2015robustness}, we adopt a robustness approach and assume that the principal designs a contract to maximize her expected payoff guarantee, given her initial knowledge.

An alternative interpretation of our model is one where a cautious agent (an entrepreneur) designs a security to sell to the principal (an investor). The agent is liquidity constrained and seeks to maximize the sale price of the security, which amounts to maximizing the principal's payoff. If the agent is cautious with respect to his higher-order beliefs---that is, his assessment of the principal's beliefs about his alternatives---then a natural objective is to design a security that maximizes his sale price guarantee. This problem coincides with our original interpretation.

A cornerstone result of our analysis is that a \textit{debt contract} is robustly optimal for the principal. Under a debt contract, the principal collects the entire prize if it falls below a predetermined debt level, while the agent keeps all revenues in excess of the debt level. Debt is optimal because it preserves the agent's option value of continued search. By promising payments to the agent only after he reaches a certain milestone, the principal discourages him from ``settling down for less'' and terminating search after exploring a cheap, safe alternative. On a technical level, debt prevails because it mirrors the Weitzman index, inducing the agent to make efficient search decisions in sequencing risky versus safe projects.

Building on the optimal debt contract, Theorem \ref{thm:optimaldebtadjacent} characterizes the set of robustly optimal contracts, which are all debt-like. We establish that a contract is robustly optimal if and only if it involves a minimum debt level and enables the principal to capture the full social surplus in the worst-case scenario. The minimum debt level is crucial to discourage the agent from terminating search too safely, while the full surplus extraction condition maximizes the principal's profit subject to the agent's participation. Our characterization encompasses several commonly observed contract formats, including pure debt, debt-plus-equity, and capped-earnout debt. These contracts strictly outperform linear (equity) contracts that are typically optimal in the robust contracting literature (e.g., \citealp{carroll2015robustness}; \citealp{walton2022general}) because linear contracts excessively dampen the agent's search incentive.

We expand on these insights in \cref{sec:extension} by extending the model in two directions. 
First, we show that when the principal initially knows multiple projects available to the agent, the qualitative conclusions of Theorem \ref{thm:optimaldebtadjacent} continue to hold. Robustly optimal contracts must feature two debt-like properties that coincide in the baseline model: a \textit{full priority} region for low outputs and a \textit{minimum claim} for the principal at high outputs. In the full priority region, the agent's wage is zero, and the principal takes the entire prize. The minimum claim property requires the principal to retain at least a certain amount of the prize for high outputs.\footnote{The first condition rules out linear (equity) contracts while the second rules out the live-or-die contracts of \citet{innes1990limited}, for example.} In the special case where the agent can repeatedly resample known projects, pure debt is uniquely optimal. 
Our second extension considers a risk-averse agent and shows that a capped-earnout contract is uniquely optimal.
Under such a contract, the agent earns wages in excess of the debt level, subject to a wage cap.
This contract balances incentivizing exploration through the debt-like structure with limiting windfall payments when the risk-averse agent's marginal benefit is low.

Our results provide a novel rationale for the prevalence of debt-like instruments arising from the interaction between search and robustness. Without employing the robust approach, optimal contracts for search reflect statistical properties of the agent's environment and need not resemble debt.\footnote{We continue this discussion in \cref{sec:assumpdiscuss}, and analyze one such model in Online Appendix~\ref{apx:bayesian}.} Without specifying that the agent searches, robustness alone delivers equity \citep{carroll2015robustness}. In that static problem, the principal guards against excessive risk-taking. Incorporating search reverses the principal's concern. In our setting, the downside of unknown risky prospects is mitigated because the agent has the option to continue exploring after a poor realization. Instead, the principal must guard against valuable exploration being crowded out by excessively safe alternatives. To discourage this inefficient search behavior, the principal withholds payments for low realizations. This provides a structural motivation for right-skewed incentives derived only from the option value inherent in the agent's decision problem.

The forces we identify arise in many environments involving delegated search beyond conventional R\&D. In \cref{sec:applications}, we discuss how our model maps into two observed contracting arrangements: earnout contracts in startup acquisitions and advance-against-royalty contracts in intellectual property development.
These applications naturally feature exploration among alternatives that may be difficult for an investor to anticipate.
We therefore complement existing explanations based on valuation uncertainty, measurement issues, or effort incentives by highlighting how these contracts can guide the agent's choice among risky alternatives.

%----------------------------------------%

\paragraph{Related Literature}

We contribute to the literature on delegated search by introducing selection among unknown projects as a new source of incentive conflict. 
Our agent chooses \textit{which project to inspect next} in addition to \textit{when to stop}. 
Existing work largely studies the duration or intensity of the agent's search, typically focusing on settings where the agent experiments with a single risky technology \citep{bergemann1998venture, bergemann2005financing, lewis2008search,  lewis2012theory}.\footnote{Many other papers on delegated search and experimentation share this emphasis and modeling choice, including \citet{horner2013incentives, manso2011motivating, zorc2025when, li2026dynamics}.}
A few models incorporate uncertainty about the agent's ability to experiment \citep{gerardi2012testing, halac2016optimal, guo_dynamic_2016, khalil2026rewarding}; closest to us, \citet{ulbricht2016optimal} studies contracting in a \citet{mccall1970economics} setup where the agent has private information about a single risky project that can be resampled. 
To instead study the agent's directed search among heterogeneous alternatives, we adopt the classic \citet{weitzman_optimal_1979} framework. 
This project selection margin connects us to a literature on delegated project choice \citep{armstrong2010model, nocke2013merger, guo2023regret}.
Our key departure from these papers is the agent's costly sequential inspection of alternatives.
To our knowledge, we are the first to study contracting for search \textit{\`a la} \citet{weitzman_optimal_1979} when the principal is uncertain about the agent's available alternatives.\footnote{\citet{postl2004delegated} studies delegated information acquisition over two alternatives. In a contemporaneous paper, \citet{hoefer2025designing} study the computational complexity of optimal contracts when the principal can contract on project identity.}

Our paper provides another setting in which a robust design approach can deliver simple, qualitative predictions. We depart minimally from the seminal model of \citet{carroll2015robustness} by incorporating search, yet our conclusion is quite different. We thus contribute to a nascent theoretical literature investigating the sensitivity of robust mechanisms to perturbations in the ambiguity set (\citealp{walton2022general}; \citealp{kambhampati_proper_2024}; \citealp{ball2024robust}; \citealp{Olszewski2025Evaluating}). Few other papers study robustness in a dynamic setting.\footnote{We direct the reader to \citet{carroll2019robustness} for a survey of this literature, which discusses some of the challenges in studying dynamics in robust contracting.} \citet{chassang2013calibrated} and \citet{liu2022robust} consider problems in which payoffs are time-separable and show that linear contracts perform well. \citet{libgober2021informational} study a durable good monopoly and show that the optimal mechanism is insensitive to the arrival of intermediate information. \citet{koh2024robust} study a learning setting similar to ours. A key difference is that they assume learning benefits the agent but may impose a negative externality on the principal, while in our model, the prize accrues to the principal and costs are borne by the agent. 

Finally, we contribute to a large finance-oriented literature that rationalizes debt as the optimal solution to different agency problems \citep{jensen1976theory}.\footnote{Seminal contributions include the costly verification models of \citet{townsend1979optimal} and \citet{gale1985incentive}. Beyond traditional agency frictions, many additional theories justify the widespread use of debt. These include lowering firms' tax burden \citep{miller1977debt}, signaling information about profitability \citep{myers1984corporate}, anticipating enforcement costs \citep{krasa200Oenforcement}, and maintaining control \citep{aghion1992incomplete,hart1994theory}. See \citet{tirole2010theory} for an overview of the literature on security design and capital structure.} In moral hazard settings, existing results showing the optimality of debt rely on statistical properties of the mapping between incremental effort and output, or restrictions to the contract space \citep{innes1990limited, poblete2012incentivecontracts, antic2014contracting, poblete2017managing, hebert2018moral}. 
Our agent's search problem is quite different from these static ones, and we impose no restrictions on the relationship between projects' costs and prize distributions, nor do we restrict attention to doubly monotone contracts. 
Debt also arises in many environments where adverse selection is central because debt is the least informationally sensitive security \citep{dang2013information}, minimizing the advantage of an exogenously informed agent.\footnote{See, for example, \citet{nachman1994optimal, demarzo1999liquidity, vanasco2017screening, yang2020optimality, malenko2020asymmetric}.}
In our setting, debt is least sensitive to the unknown projects available to the agent in the endogenous worst-case scenario.

%----------------------------------------%
%----------------------------------------%
\section{Model}
\label{sec:model}

A principal (``she'') contracts with an agent (``he''), who sequentially searches through projects to generate a prize. Each project is a ``Pandora's box'' in the language of \cite{weitzman_optimal_1979}, and the set of projects available to the agent is denoted by $\mathcal{A} = \left\{ a_i \right\}_{i=0}^n$, with $n \in \mathbb{N}$.  The project $a_i$ is described by a pair $(F_i, c_i) \in \Delta(\mathbb{R}_+) \times \mathbb{R}_+ $ such that $F_i$ has finite expectation. The parameter $c_i$ represents the agent's private \textit{cost} to learn project $i$'s realized \textit{prize} $y_i$, which is distributed according to $F_i$. Each project's prize is drawn independently, and we assume both players are risk-neutral.

When facing a set of projects $\mathcal{A}$, the agent engages in sequential search (with recall) \textit{\`a la} \citet{weitzman_optimal_1979}. We can describe the agent's strategy as a function of two state variables: (i) the set of projects sampled to date, denoted by $\tilde{\mathcal{A}} \subseteq \mathcal{A}$, and (ii) the highest monetary wage from the sampled projects, denoted by $\tilde y$. Formally, a strategy is a function $\sigma: 2^\mathcal{A} \times \mathbb{R} \rightarrow {\mathcal{A}} \cup \{\emptyset\}$, where $\sigma(\tilde{\mathcal{A}}, \tilde y) = a_i$ means that the agent will sample project $a_i \in \mathcal{A}\setminus\tilde{\mathcal{A}}$ next and $\sigma(\tilde{\mathcal{A}}, \tilde y) = \emptyset$ means that he will cease searching and present a prize that gives him the monetary reward of $\tilde y$. The agent may always return a prize of zero without sampling any projects. We abusively write $a_i \in \sigma$ to denote the event that the project $a_i$ is sampled according to the strategy $\sigma$. 

As in a typical moral hazard environment, we assume that the agent's action---i.e., the search process---is unobservable and cannot be contracted upon. In particular, the principal does not know which projects the agent sampled to generate a prize. However, she can contract on the final prize $y$ the agent presents to her.\footnote{We assume the agent can only present one \emph{single} prize, as is often the case in practice. In \Cref{sec:assumpdiscuss}, we argue that the principal does not gain from offering general mechanisms, including allowing the agent to present more than one prize.} The principal's only incentive tool is thus a wage contract $w: \mathbb{R}_+ \rightarrow \mathbb{R}$, where $w(y)$ is the agent's monetary payment when the presented prize is $y$. All contracts must be measurable and satisfy limited liability, meaning $w(y)\geq 0$ for any $y\in \mathbb{R}_+$.

The timing of the game is summarized as follows.
\begin{enumerate}
    \item The principal sets a contract $w$.
    \item The agent sequentially searches among $\mathcal{A}$, after which he presents a prize $y$. 
    \item The agent's payoff is $w(y) - \sum_i c_i \mathbf{1}_{[a_i \in \sigma]}$ and the principal's payoff is $y-w(y)$.
\end{enumerate}

Although the agent knows the true set of projects $\mathcal{A}$, we assume the principal knows only one of those projects, $a_0 = (F_0, c_0)$, at the time of contracting. So, she considers all $\mathcal{A} \supseteq \mathcal{A}_0:= \{a_0\}$ as possible. We assume that the known project generates a positive surplus: $\mathbb{E}_{F_0}[y]>c_0$. 

The principal's objective is to determine a wage contract $w$ that maximizes the worst-case expected payoff against all possible realizations of $\mathcal{A}$ consistent with $\mathcal{A}_0$. Formally, given a realized set of projects $\mathcal{A}$ and a wage contract $w$, we let $\Sigma(w,\mathcal{A})$ denote the set of optimal search strategies for the agent. For a given strategy $\sigma$, we write $\mathbb{E}_{\sigma}$ to denote the expectation with respect to the induced distribution over prizes generated by the agent's search. The principal's payoff for a given contract and set of projects is
\begin{equation}
    V_P(w \mid\mathcal{A}):= \sup_{\sigma \in \Sigma(w,\mathcal{A})} \mathbb{E}_\sigma[y-w(y)], \nonumber
\end{equation}
where we assume the agent breaks ties in favor of the principal.\footnote{This assumption is for ease of exposition, and is not essential to the results. If we instead assumed adversarial tie-breaking, a perturbation of the contracts we discuss would lead to an arbitrarily small loss in profit while giving the agent strict incentives.}  

The principal evaluates a wage contract $w$ by its payoff guarantee $V_P(w)$, her expected payoff in the worst-case realization of $\mathcal{A}$: 
\begin{equation}
V_P(w):=\inf_{\mathcal{A} \supseteq \mathcal{A}_0} V_P(w \mid\mathcal{A}) .  \nonumber
\end{equation}

The principal seeks a contract that maximizes her payoff guarantee $V_P(w)$. We denote the optimal payoff guarantee by  $V_P$. We call a contract that achieves a guarantee of $V_P$ a (robustly) optimal contract.
\begin{equation}
    V_P := \sup_w V_P(w)\nonumber
\end{equation}

%----------------------------------------%
%----------------------------------------%
\section{Optimal Contracts}
\label{sec:optimal}
%----------------------------------------%
    \subsection{Agent's Response to a Contract}
    \label{subsec:agentstrategy}
        
        Our main objective in \Cref{sec:optimal} is to characterize all of the principal's robustly optimal contracts. To do so, we begin by describing the agent's best response to a contract. 
        
        The agent's optimal search strategy depends on the offered contract and the available set of projects. As a first step, we recall the solution to the problem of \citet{weitzman_optimal_1979}. For the project $a_i = (F_i, c_i)$, we define its \textit{index} (or \textit{reservation value}),  $r_i$, as the smallest solution to\footnote{When $c_i$ is positive, the solution to \cref{eq:weitzmanindex} is unique. However, when $c_i = 0$, there may be a continuum of solutions to \cref{eq:weitzmanindex}, including infinity. The index may also be negative, in which case it is never optimal for the agent to sample the project.}
        \begin{equation}
        \label{eq:weitzmanindex}
            c_i = \int [y_i - r_i]^+ dF_i(y_i). 
        \end{equation} 
        A social planner who maximizes the joint welfare of the principal and the agent facing the same search problem will sample the projects in descending order of their index values and conclude the search whenever a realized prize $y$ is larger than the index values of the remaining unsampled projects.\footnote{If there are multiple projects with the same index, the agent is indifferent to their ordering, and there are multiple optimal strategies.}
        
        When facing the wage contract $w$, the agent still uses an index strategy, except his incentive is distorted by the contract. Given a wage contract $w$, project $a_i$'s \textit{$w$-induced index} (or \textit{$w$-induced reservation value}), $r_i^w$, is the smallest solution to
        \begin{equation}
        \label{eq:w_weitzmanindex}
            c_i = \int [w(y_i) - r_i^w]^+ dF_i(y_i). 
        \end{equation} 
        Thus, the agent optimally samples projects in descending order of $r_i^w$ and stops searching when his best realized monetary payoff exceeds the $w$-induced index values of the remaining unsampled projects.
        
%----------------------------------------%   
    \subsection{Characterization of Optimal Contracts}
    \label{subsec:optimalcontracts}
        
        Having specified the agent's optimal response to any contract $w$, we now return to the principal's problem of finding the contract that maximizes the payoff guarantee. We begin by identifying a simple upper bound on this payoff guarantee: the surplus of the known project $a_0$, which we denote by $s_0$.
        \begin{equation*}
            s_0 := \mathbb{E}_{F_0}[y] - c_0
        \end{equation*}
        
        \begin{observation}
        \label{obs:ub}
        No contract can guarantee the principal more than the full surplus of the known project.  That is, for all $w$, 
        \begin{equation}
        \label{eq:ub}
            V_P(w) \leq V_P(w \mid \mathcal{A}_0) \leq s_0 .
        \end{equation}
        
        \end{observation}

        The first inequality follows because the principal's guarantee is bounded above by her profit when the agent has no unknown projects. The second follows from the agent's participation constraint. When the agent actually has no unknown projects, the principal's profit is at most $s_0$; any greater and the agent's payoff from search would be negative, so he would rather produce zero.         
        
        It is not immediately clear that this upper bound is useful, as robustly extracting the entire known surplus is a demanding desideratum. Nonetheless, our next proposition shows that a properly calibrated \textit{debt contract} can attain it. 
     
        For any $z \geq 0$, the \textbf{$z$-debt contract} is defined as the contract $w(y) := [y-z]^+$. When the principal offers the $z$-debt contract, she collects all the returns up to the debt level $z$, after which the agent becomes the residual claimant and collects any prize in excess of $z$. Conceptually, this is equivalent to giving the agent a call option on the prize with a strike price of $z$. 
        
        Let $r_0$ denote the index of the known project $a_0$ as defined using \Cref{eq:weitzmanindex}, and let $w_0$ denote the $r_0$-debt contract, $w_0(y) = [y - r_0]^+$.

        \begin{proposition}
        \label{thm:debtcontracts}
        The payoff guarantee of the $r_0$-debt contract, $w_0$, is the entire surplus $s_0$. Therefore, $w_0$ is robustly optimal, and $V_P = V_P(w_0) = s_0$.
        \end{proposition}
        \smallskip 

        \begin{proof}
        Given the upper bound identified in \Cref{obs:ub}, it suffices to show $V_P(w_0) = s_0$. 
        
        First, we argue that $V_P(w_0) = V_P(w_0 \mid \mathcal{A}_0)$, i.e., the worst-case scenario is $\mathcal{A} = \mathcal{A}_0$ \emph{when the principal offers $w_0$}. Fix any arbitrary $\mathcal{A} \neq \mathcal{A}_0$. We distinguish two cases by whether the agent samples $a_0$ for a given realization of prizes $\{y_i\}_{i=0}^{n}$. Consider first the case where the agent \textit{does not} sample $a_0$ before stopping. Given that $r_0^{w_0} = 0$, the agent must have obtained a positive wage from another prize and terminated search.\footnote{In the knife-edge case when the agent gets a prize of exactly $r_0$, both he and the principal are indifferent between continuing and stopping. We assume he continues to search $a_0$ in this case.} 
        
        This happens under the contract $w_0$ only when his presented prize is higher than $r_0$, which implies that the principal has already fully recouped the debt. Notice that $r_0 \geq s_0$ because $r_0 - s_0 = \int [r_0 - y]^+ dF_0(y) \geq 0$, so this outcome is better for the principal than the case when only $a_0$ is available. Now consider the second case where the agent \textit{does} sample $a_0$. Because both $w_0(y)$ and $y-w_0(y)$ are weakly increasing in $y$, any projects searched in addition to $a_0$ can only improve the distribution of the returned prize in the sense of first-order stochastic dominance. In both cases, the principal is weakly better off under $\mathcal{A}$ relative to $\mathcal{A}_0$. 

        Second, we show that $V_P(w_0 \mid \mathcal{A}_0) = s_0$ because the calibrated debt level $r_0$ leaves the agent zero expected payoff when $\mathcal{A}=\mathcal{A}_0$. To see this, notice that 
        \begin{equation}\label{eq:debtnosurplus}
            \mathbb{E}_{F_0}[w_0(y)] - c_0 = \mathbb{E}_{F_0}[(y-r_0)^+] - c_0 = 0, \nonumber
        \end{equation}
        where the last equality follows from the definition of the index $r_0$. In other words, when $a_0$ is the only available project, the contract $w_0$ maximizes total surplus (because the agent is still willing to sample $a_0$) while leaving no surplus to the agent. This also implies that the $w_0$-induced index of the known project $a_0$ is $r_0^{w_0} = 0$. 

        Therefore, $V_P(w_0) = V_P(w_0 \mid \mathcal{A}_0) = s_0$, and thus $w_0$ is robustly optimal.
        \end{proof}

        As the proof demonstrates, $w_0$ attains the payoff guarantee upper bound because it achieves two goals simultaneously. First, $w_0$ guards the principal against the agent's early termination of search. Because the agent is not rewarded for presenting low prizes, any prize that can terminate his search before exploring $a_0$ must be sufficiently large. As a result, when offering $w_0$, having additional projects beyond $a_0$ always (weakly) benefits the principal. Second, in the worst case where $a_0$ is the only available project, the debt level $r_0$ is chosen so that the principal extracts the entire surplus. Together, these two properties maximize the principal's payoff guarantee.
        
        %--------------------------------------------------------------------------------%
        %--------------------------------------------------------------------------------%

        It turns out that debt contracts are not the only contracts that can achieve this upper bound. Following the preceding argument, any contract that causes both inequalities in \Cref{eq:ub} to bind will achieve the same guarantee. Satisfying the first equality, $V_P(w) = V_P(w \mid \mathcal{A}_0)$, requires a contract that prevents the known project from being crowded out by a low-value alternative, constraining the shape of the contract. The second equality, $V_P(w \mid \mathcal{A}_0) = s_0$, pins down the level of the contract to extract the full surplus of the known project $a_0$. These conditions are also necessary for a contract to deliver a guarantee of $s_0$. Using this logic, the following theorem characterizes the properties of robustly optimal contracts. 
        
        \begin{theorem}
        \label{thm:optimaldebtadjacent}
        A contract $w$ is robustly optimal if and only if it satisfies the following two conditions. 
            \begin{enumerate}
                \item \textbf{Minimum Debt Level condition} (MDL henceforth): \\
                $$w(y) \leq [y-s_0]^+$$ 
        
                \item \textbf{Full Surplus Extraction condition} (FSE henceforth):\\
                $$\mathbb{E}_{F_0}[w(y)] = c_0.$$
            \end{enumerate}
        \end{theorem}
        \begin{proof}
            See Appendix~\ref{proof:thm:optimaldebtadjacent}.
        \end{proof}

        The intuition for the ``if'' part of this theorem parallels that of \Cref{thm:debtcontracts}. The MDL condition guarantees that $\mathcal{A} = \mathcal{A}_0$ is indeed the worst case; that is, having additional projects beyond $a_0$ never harms the principal. As before, the agent is paid nothing if the presented prize falls below $s_0$, whereas the principal's payoff $y-w(y)$ is at least $s_0$ if the presented prize exceeds $s_0$.\footnote{Notice that the debt level is not necessarily as high as $r_0$, which was the level of the pure debt contract in \cref{thm:debtcontracts}.}  This guards the principal against unknown projects crowding out $a_0$ because the principal will at least collect $s_0$ even if the agent terminates the search before sampling $a_0$. At the same time, the FSE condition ensures that the principal collects the full surplus when $a_0$ is the only available project.
        
        The ``only if'' part is shown by construction. From \Cref{eq:ub}, any contract $w$ with a profit guarantee of $s_0$ must have $V_P(w \mid \mathcal{A}_0) = s_0$, which means FSE must hold. This implies that the $w$-induced index of the known project $a_0$ satisfies $r_0^w = 0$. Then, suppose MDL is violated, i.e., there exists $y'$ such that $w(y') > [y'-s_0]^+$. In this case, a cheap and safe project $a_1 = (\delta_{y'}, 0)$, where $\delta_{y'}$ is a Dirac mass at $y'$, can crowd out $a_0$ because it gives the agent a strictly positive wage while $r_0^w = 0$. In other words, when $\mathcal{A}=\{a_0,a_1\}$, the agent will only explore $a_1$ and stop. This adversarial scenario yields the principal a payoff strictly lower than $s_0$, violating the optimality of $w$. 

        The MDL condition in \cref{thm:optimaldebtadjacent} embeds two related implications, which we later show generalize to the case where the principal knows multiple projects available to the agent. 
        The first is \textit{principal priority} for low outputs (below $s_0$). In this region, the agent is paid nothing. By requiring the agent to reach a certain standard to receive payment, the principal discourages the agent from ``settling down'' for cheap, safe alternatives. That is, the contract incentivizes the agent to explore risky alternatives.
        The second implication is a \textit{minimum claim} for high outputs (above $s_0$). For large prizes, the principal's ex-post profit is at least $s_0$, preventing the principal from overpaying the agent when unexpectedly successful outcomes arise. Under the MDL condition, the principal priority and minimum claim thresholds coincide at $s_0$; however, we show in \cref{sec:extension} that these thresholds may differ when the principal knows more about the agent's projects.

        The set of robustly optimal contracts identified in \Cref{thm:optimaldebtadjacent} includes several types of contracts that are commonly observed in practice. We depict three prominent examples in \Cref{fig:alloptimalcontracts}.  The first type of contract is \textit{pure debt} (previously, just called ``debt''), in which the agent becomes the full residual claimant after repaying the debt. 
        The second type of contract is \textit{debt-plus-equity}. Under such a contract, the principal collects all the returns up to the debt level, after which the agent is a partial residual claimant. In practice, this can be implemented as a convertible debt, a contract that gives the principal the option to convert her debt into equity participation at a pre-specified rate. 
        The third type of contract is \textit{capped-earnout debt}, which is similar to a debt contract except that the agent's wage (or ``earnout payment'') is bounded above. This contract resembles a senior tranche used in structured finance. Such a contract makes the agent fully sensitive to marginal changes in the prize for intermediate levels of the prize, but insensitive outside this range. 
        
        In \cref{sec:extension}, we show that enriching the model delivers unique predictions among these contracts. We then map our model and predictions into several contracts used in practice in \Cref{sec:applications}. 
    
    \begin{singlespace}
    \begin{figure}[htbp]
        \begin{center}
    	\subfigure[Pure debt]{
    		\begin{minipage}[t]{0.3\textwidth}
    			\centering
    		\begin{tikzpicture}[xscale=0.36, yscale = 0.36]
    			\draw [thick, <->] (0,10) -- (0,0) -- (10,0);
    \draw [dashed] (0,0) -- (9,9);
    \begin{scriptsize}
                \node [right] at (10,0) {$y$};
                \draw [thick, red] (0,0) -- (4.5,0) -- (10,5.5) node [right] {$w_0(y)$};
    			\draw [thick, blue] (0,0) -- (4.5,4.5) -- (10,4.5) node [right] {$y-w_0(y)$};
    			\draw [-, thick] (4.5,0.1) -- (4.5,-0.15) node [below] {$r_0$};
    			\draw [-, thick] (2,0.1) -- (2,-0.1);
                \node [below] at (1.7,-0.1) {$s_0$};
    \end{scriptsize}
            \end{tikzpicture}    
    		\end{minipage}
    	}
    	\subfigure[Debt-plus-equity]{
    		\begin{minipage}[t]{0.3\textwidth}
    			\centering
    		\begin{tikzpicture}[xscale=0.36, yscale = 0.36]
    			\draw [thick, <->] (0,10) -- (0,0) -- (10,0);
    \draw [dashed] (0,0) -- (9,9);
    \begin{scriptsize}
                \node [right] at (10,0) {$y$};
    			\draw [thick, red] (0,0) -- (3,0) -- (10,4.2) node [right] {$w(y)$};
    			\draw [thick, blue] (0,0) -- (3,3) -- (10,5.8) node [right] {$y-w(y)$};
    			\draw [-, thick] (2,0.1) -- (2,-0.1);
                \node [below] at (1.7,-0.1) {$s_0$};
    \end{scriptsize}
    
    		\end{tikzpicture}    
    		\end{minipage}
    	}
    	\subfigure[Capped-earnout debt]{
    		\begin{minipage}[t]{0.3\textwidth}
    			\centering
    		\begin{tikzpicture}[xscale=0.36, yscale = 0.36]
    			\draw [thick, <->] (0,10) -- (0,0) -- (10,0);
    \draw [dashed] (0,0) -- (9,9);
    \begin{scriptsize}
                \node [right] at (10,0) {$y$};
    			\draw [dotted] (4.7,2.0) -- (0,2.0);
    			\draw [-, thick] (2,0.1) -- (2,-0.1);
                \node [below] at (1.7,-0.1) {$s_0$};
    			\draw [thick, red] (0,0) -- (2.7,0) -- (4.7,2.0) -- (10,2.0) node [right] {$w(y)$};
    			\draw [thick, blue] (0,0) -- (2.7,2.7) -- (4.7,2.7) -- (10,8) node [right] {$y-w(y)$};
    \end{scriptsize}
    
    		\end{tikzpicture}    
    		\end{minipage}
    	}
        \end{center}
    \caption{Three examples of optimal contracts. Panel (a) depicts the pure debt contract $w_0$. Panel (b) depicts a debt-plus-equity contract. Panel (c) depicts a capped-earnout debt contract. In each panel, the red and the blue curves represent the agent's and the principal's payoffs, respectively.}
    \label{fig:alloptimalcontracts}
    \end{figure}
    \end{singlespace}

    \paragraph{Why not linear contracts?}

    Our model departs from the seminal model of \citet{carroll2015robustness} in one aspect only: rather than choosing a project as a one-time decision, the agent in our framework sequentially searches across alternatives. This difference leads to qualitatively different conclusions. Whereas much existing robust literature finds the optimality of linear (equity) contracts, we find debt.\footnote{Papers that show optimality of linear contracts in robust settings include \citet{hurwicz1978incentive}, \citet{carroll2015robustness}; \citet{walton2022general}; \citet{dai2022robust}; \citet{liu2022robust}, and \citet{kambhampati_randomization_2025}.} We propose two lenses through which to understand this distinction. 
    
    First, our principal guards against a fundamentally different type of unknown project than the principal in \citet{carroll2015robustness}. In \citet{carroll2015robustness}, the principal worries that the agent selects an excessively risky project and gambles for a favorable realization. Such unknown risky projects are a boon in our setting. If the gamble returns an unfavorable realization, the agent can abandon the outcome and continue exploring the next project. Our principal does not need to protect herself against risky prospects, but aims to prevent the agent from inefficiently initiating search with a safe, low-value project and then terminating his search. Indeed, this is the essence of the ``only if'' argument following \cref{thm:optimaldebtadjacent}. If the principal tried to extract the full surplus using a linear contract instead of debt, the payoff guarantee would be zero. In this case, the known project could be crowded out by a free, safe project with an arbitrarily low prize. Thus, debt contracts offer a distinct advantage because they make low prizes unattractive to the agent, thereby encouraging risk-taking and continued exploration.
    
    Second, the robustly optimal contracts in the one-shot versus sequential search models reflect different ``sufficient statistics'' of a project that guide the agent's choice in each environment. In \citet{carroll2015robustness}, the social value of a project is captured by its \textit{expected} prize net of cost. This makes linear contracts desirable, since they align the agent's and principal's incentives over different risk distributions in the static setting. In our dynamic search setting, however, the relevant statistic is the \textit{Weitzman index}, which captures the option value of continued search. A pure debt contract arises as optimal because it mirrors the index: both evaluate the expected prize in excess of a threshold. As a result, under a debt contract, our agent's ranking of safe versus risky projects coincides with the efficient one. This leaves less room for a malevolent nature to interfere, as any unknown projects can only improve social surplus and simultaneously the principal's payoff.

%----------------------------------------%   
    \subsection{Discussion of Assumptions}
    \label{sec:assumpdiscuss}
    
        \paragraph{Robustness.}

        We employ the robust criterion as a way to capture the uncertainty inherent in financing innovation. This approach is especially relevant when the principal may have difficulty forming a credible prior, either because she lacks the agent's expertise or because it may be difficult to anticipate new opportunities that arise after the contract is written. That said, the max-min approach is deliberately conservative. If, instead, the principal had more reliable information about the agent's alternatives or put more weight on upside performance, other considerations beyond those in our model would naturally become relevant.

        We analyze a simple Bayesian benchmark of our model in Online Appendix~\ref{apx:bayesian} to illustrate this comparison.
        Motivated by the worst-case scenarios we identify, we consider the agent's search over a commonly known risky alternative and a privately known safe alternative. 
        This benchmark highlights a familiar tradeoff for the principal: she wants to preserve search efficiency while economizing on the agent's rents. 
        The first force pushes towards right-skewed incentives, while the second depends on statistical features of the agent's problem.
        This may lead the agent's search to be inefficient, with wages reflecting the informativeness of the output about the agent's search behavior \citep{holmstrom1979moral}.
        We show that the optimal doubly monotone contract exhibits a minimum debt level. Roughly, this is because continued search is most valuable when the safe alternative generates a low prize. Withholding payments for low prizes maintains this incentive while reducing the agent's rents.
        Under an additional distributional condition provided in the appendix, the right-loading of incentives for efficiency reasons coincides with rent-minimization, and pure debt becomes optimal.

        Our robust model provides complementary insights. 
        Beyond offering a tractable way to study a large, unstructured uncertainty space, robustness identifies a mechanism and contractual prescription that are independent of the statistical details of the agent's alternatives. 
        Debt arises because it preserves the agent's ranking of safe and risky projects, preventing known projects from being crowded out.
        This provides a structural justification for debt arising from the option value in the agent's problem regardless of how likely certain alternatives may be.
        The Bayesian model we study preserves this force, while showing how it can be reinforced or offset by distribution-specific information.      
    
        \paragraph{Contracting space, screening, and randomization.}

        Following the logic of \cref{obs:ub}, any mechanism that respects the agent's participation cannot outperform the debt contract's payoff guarantee of $s_0$. This includes screening, randomization, or contracting on the project identity. For instance, consider the class of multi-stage mechanisms of \citet{myerson1982optimal} in which the agent reports his private information (the set of projects), is given an action recommendation, and the agent's action generates contractible signals. Consider any mechanism $M$ satisfying interim participation, i.e., for every $\mathcal{A} \supseteq \mathcal{A}_0$, there is some strategy delivering non-negative expected payoff to the agent. For this mechanism, we must have $V_P(M \mid \mathcal{A}_0) \leq s_0$ because $s_0$ is the total surplus available when $\mathcal{A}=\mathcal{A}_0$. Since $V_P(M) \leq V_P(M \mid \mathcal{A}_0)$, no mechanism can have a strictly higher guarantee than contracting on output alone. 
        
        However, more complex contracting mechanisms may weakly dominate offering the simple debt contract. For instance, suppose the principal could verify the parameters of the box that yielded the agent's prize. By paying the agent with a debt contract pegged to the index of the returned project, the principal could robustly extract the full expected surplus of \textit{any} set of projects.

        \paragraph{Limited liability.} 

        Limited liability plays a slightly different role in our model than in the security design literature at large. As is typical, if we allow the agent’s wage to be negative, the principal can ``sell the firm’’ to the agent for the expected surplus, using the contract $w(y) = y - s_0$. Thus, limited liability is clearly important in the necessity direction of \Cref{thm:optimaldebtadjacent}, where we characterize all optimal contracts. However, relaxing limited liability does not interfere with sufficiency because the contracts we identify give the same worst-case payoff as selling the firm, so they remain robustly optimal. 
        
        Moreover, selling the firm to the agent is weakly dominated by any of the contracts identified in \Cref{thm:optimaldebtadjacent}. Intuitively, selling the firm must yield a deterministic payoff of $s_0$---the sale price---regardless of the true set of projects $\mathcal{A}$. By contrast, for any robustly optimal contract $w$ identified in \Cref{thm:optimaldebtadjacent}, the principal's expected payoff is weakly greater than $s_0$ across all realizations of $\mathcal{A}$, and strictly greater for some. 

        \paragraph{Refinements.}

         Within our baseline model, there is no obvious criterion to select among the contracts identified by \Cref{thm:optimaldebtadjacent}. Notably, all are undominated; each delivers the same payoff guarantee and is the principal's preferred contract for some realization of $\mathcal{A}$. An alternative refinement based on the total surplus generated by a contract is similarly inapplicable. Unfortunately, in general, no contract maximizes the total surplus against every $\mathcal{A} \supseteq \mathcal{A}_0$ while remaining robustly optimal.

%----------------------------------------%
%----------------------------------------%
\section{Extensions}
\label{sec:extension}
%----------------------------------------%
    \subsection{Multiple Known Projects}
    \label{subsec:manybox}
    
    Although our baseline model assumes that the principal knows only one of the agent's projects, in some settings, it may be more natural for the principal to know multiple available alternatives at the outset. In this extension, we show that qualitative features of robustly optimal contracts in the baseline model---principal priority for low presented prizes and minimum claim for high presented prizes---continue to hold when $\mathcal{A}_0$ is non-singleton. Formally, we depart from the baseline model by letting the principal know $k$ projects available to the agent, so $\mathcal{A}_0 = \{a_0, a_1, ..., a_{k-1}\}$. We continue to let $V_P$ denote the principal's optimal payoff guarantee in this setting. 
   
    \begin{proposition}
    \label{prop:twobox}
    Suppose $\mathcal{A}_0$ is non-singleton. For any robustly optimal contract $w$, there exist two thresholds $0 < \underline{y} \leq \overline{y} < \infty$ satisfying
    \begin{itemize}
        \item[(a)] Principal priority: $w(y) = 0$ for all $y \leq \underline{y}$,
        \item[(b)] Minimum claim: $w(y) \leq y-V_P$ for all $y \geq \overline{y}$.
    \end{itemize}    
    \end{proposition}
    \begin{proof} 
    See Appendix~\ref{pf:prop:twobox}. 
    \end{proof}
    
    The intuition behind principal priority is as follows. Given an optimal contract $w$, there must be a known project $a_i\in\mathcal{A}_0$ that (i) the principal finds ``valuable'' in the sense that she wants the agent to explore it on some equilibrium path and (ii) has zero $w$-induced index. If not, adding some debt to the contract extracts more surplus from the agent without affecting his search behavior, thus increasing the principal's guarantee. For a project with zero $w$-induced index to remain valuable to the principal in the worst case, the agent's wage must be zero for sufficiently low realizations. Otherwise, this zero-index project has no value, as it can be crowded out by a free project with arbitrarily low output. Thus, there must be a region where the agent is paid zero to ensure profits from the lowest induced index project. 
    
    The minimum claim property follows from reasoning similar to that in the baseline model. If the agent is presented with a project that delivers a constant wage higher than the maximal $w$-induced index among the known projects, the agent will take this safe project and terminate search. The minimum claim property ensures that the principal's payoff is at least $V_P$ in this case. 
    
    This proposition suggests that the baseline model's qualitative insights do not critically depend on the assumption that the principal knows only one of the agent's projects. As in the baseline model, the principal uses debt-like incentives to guard against valuable projects being crowded out. That said, the baseline model yields a neat characterization of all optimal contracts because we can pin down $V_P$. When the principal knows more than one project, a characterization is possible only in a few special cases. 
    
    One case yielding a tight characterization is when the known projects can be repeatedly sampled, meaning $\mathcal{A}_0$ contains infinite replicas of each $(F_i, c_i)$ pair. In this specification, pure debt is the essentially unique robustly optimal contract. Let $r^*$ denote the maximal index of all projects in $\mathcal{A}_0$. 

    \begin{proposition}
    \label{prop:resampling}
            Suppose projects can be resampled. The pure debt contract $w^*(y) = [y - r^*]^+$ is optimal. Moreover, every optimal contract lies pointwise below $w^*$, and if there is a unique project $(F^*,c^*) \in \mathcal{A}_0$ with index $r^*$, then every optimal contract coincides with $w^*$, $F^*$-almost everywhere.
    \end{proposition}
    
    \begin{proof}
        See Appendix \ref{proof:prop:resampling}.
    \end{proof}

    When resampling is possible, the principal's payoff guarantee becomes $V_P = r^*$, which coincides with the efficient surplus. The pure debt contract achieves this bound for the same reason as in the baseline model, as the agent is incentivized to search until he clears the debt level. The fact that $w^*$ is the largest optimal contract follows from the upper bound of \cref{prop:twobox}. Finally, where there is a unique project with index $r^*$, in order to robustly extract the full surplus, the contract must satisfy $\mathbb{E}_{F^*}[w(y)] = c^*$, implying $F^*$-uniqueness. 

    %----------------------------------------%

    \subsection{Risk-Averse Agent}
    \label{subsec:riskaverse}

    Risk neutrality for both the principal and the agent contributes to the multiplicity of optimal contracts in \Cref{thm:optimaldebtadjacent} because every fully extractive contract generates the same surplus. To adjust the baseline setup, we now assume that the agent has a known, continuous, strictly increasing, and strictly concave utility function $u(\cdot): \mathbb{R}_+ \rightarrow \mathbb{R}_+$ over monetary rewards. We normalize $u(0) = 0$ and continue to assume that the principal is risk-neutral. Finally, we strengthen the positive surplus assumption to $\mathbb{E}_{F_0}\left[u(y)\right] >c_0 >0$ to simplify the exposition. 

    We say that a contract $w$ is a capped-earnout debt contract if it takes the form $w(y) = \min\{\bar{w}, [y - z]^+ \}$ for parameters $z,\bar{w} \in \mathbb{R}_+$. The parameter $z$ defines the debt level, as in the baseline setup, while $\bar{w}$ is the wage cap. 
    
    \begin{proposition}
    \label{prop:tranche}
    If the agent is risk-averse, a capped-earnout debt contract  $w^*(y) = \min\{\bar{w}, [y - z]^+ \}$ is robustly optimal. The parameters $z$ and $\bar{w}$ are jointly determined by the following two conditions:
        \begin{enumerate}
        \item \textbf{RA - Minimum Debt Level (RA-MDL):}\ $z = \mathbb{E}_{F_0}[y - w^*(y)]$ 
        \item \textbf{RA - Full Surplus Extraction (RA-FSE):}\ $\mathbb{E}_{F_0}[u(w^*(y))] = c_0$.
        \end{enumerate}
    Under $w^*$, the payoff guarantee coincides with the debt level,  $V_P = z$. Moreover, $w^*$ is $F_0$-uniquely optimal.\footnote{Without additional assumptions on $F_0$ and the shape of the agent's utility $u$, the wage cap may satisfy $\bar{w} = \infty$, i.e., the contract becomes pure debt. A sufficient condition for $\bar{w} < \infty$ is that the prize $0$ is in the support of $F_0$. More generally, whenever the known project carries meaningful risk---in that the agent could draw a prize below $r_0$---we have a finite wage cap $\bar{w} < \infty$.}
    \end{proposition}
    
    \begin{proof}
    See \Cref{proof:prop:tranche}.
    \end{proof}
    
    When the agent is risk-averse, the principal has two conflicting objectives. On the one hand, ignoring the agent's incentives, fully insuring the agent (i.e., providing a constant wage contract) is socially efficient and would allow the principal to extract the most surplus. On the other hand, to preserve the agent's search incentives, the principal needs a contract that withholds wages for low prizes. The capped-earnout debt contract balances these objectives. As in the baseline model, the principal's robustness concern means she must pay the agent zero wage if the presented prize is below $V_{P}$ and collect at least $V_{P}$ if the prize is above that. Subject to this ``debt constraint,'' the principal would like to minimize the agent's risk exposure, thereby appropriating more of the surplus. The best way to do so is to smooth the agent's wage only above a certain threshold, leading to the capped-earnout debt contract we characterize. 

    The proof formalizes this logic by showing that the capped-earnout debt contract satisfies a particular fixed-point property. Namely, the principal's expected profit when $a_0$ is the only available project exactly coincides with the profit when $a_0$ is crowded out by a safe project. No contract can simultaneously improve upon both of these profit levels, implying this is an upper bound on the principal's payoff. Since the capped-earnout debt contract achieves this bound, it must be optimal.
    
    Intuitively, the three regions of a capped-earnout debt contract serve different purposes. The initial flat region aims to discourage settling down for less. The intermediate increasing region encourages exploration as the agent gains from the upside. The final capped region limits windfall rents and protects the principal from overpaying in extremely successful states where the agent's marginal benefit is relatively low. As we argue in the next section, these motivations are fundamental to the actual implementation of earnout contracts.    
%----------------------------------------%
\section{Contracts for Search in Practice}
    \label{sec:applications}

    Across many environments featuring delegated search, we observe structurally similar contracts: the agent receives little or no compensation for low outcomes while retaining participation in successes. Our model provides one explanation for this debt-like component. Right-skewed payments preserve the agent's incentives to efficiently pursue riskier, higher-value projects. 
    
    Beyond the above general insight, we discuss two applications that closely fit the setting and predictions of the model. Before turning to them, however, we should be up-front about the limits of our analysis. Our model focuses on the agent's incentives to search and abstracts from other considerations that affect observed contracts, such as principal moral hazard,  measurement issues, timing considerations, and control rights. Accordingly, we interpret our model as isolating one force shaping the contracts described below, rather than as providing a complete theory of their design.

    \paragraph{Earnout contracts.}

    Earnout contracts are used in acquisitions as a method to make part of the acquisition price contingent on the target's future performance. 
    In practice, earnouts often appear in settings that map naturally onto our model.
    To make this mapping explicit, we view the acquirer as the ``principal'' and the retained management of the target company as the ``agent.'' 
    The target's existing product or commercialization plan is interpreted as the known project, and the principal's ambiguity set includes new technologies, applications, or cross-company synergies that were unknown at the time of contracting. 
    We treat the earnout contract as the wage contract itself, abstracting from the upfront payment typically made in acquisitions.

    Empirical work on the incidence of earnout contracts is broadly aligned with our interpretation.
    Speaking to the importance of uncertainty and the agent's experimentation, earnouts are disproportionately used in the acquisitions of private companies operating in  R\&D-intensive or intangible-rich industries where management is retained \citep{datar2001earnouts,dahlen2025earnouts,barbopoulos2021real}. 
    Our view of the post-acquisition phase featuring a moral hazard problem is strongest when the retained management has meaningful discretion over the acquired company's operations, and may therefore pursue the projects they deem fit.
    Empirically, earnout acquisitions are likely to maintain the target as a separate subsidiary, supporting a degree of managerial autonomy \citep{kohers2000earnouts, barbopoulos2021real}. 
    
    A typical earnout contract is structured so that payments increase in excess of a performance target, up to a maximum payout \citep{cain2011earnouts}. 
    Despite documented heterogeneity in details such as performance measure or term length, these contracts largely resemble the capped-earnout debt contracts derived in \cref{subsec:riskaverse}. 
    Occasionally, observed contracts \textit{exactly} implement the compensation structure predicted in \cref{prop:tranche}. We provide one such example in Appendix \ref{appendix:earnout_example}; several others are publicly retrievable from the Securities and Exchange Commission. Our analysis justifies the three regions observed in earnout contracts through the retained management's search incentives.
    The initial flat region discourages stopping with low realizations, and the increasing region above the performance target preserves some upside incentive for risky exploration. 
    The final payment cap limits windfall rents and protects the acquirer from overpaying in successful states that provided minimal ex-ante incentives. 
       
    Our results complement existing work that provides empirical explanations for why earnout contracts arise, but says relatively little about the effects of contractual design \citep{dahlen2025earnouts}. 
    Justifications for these contracts include valuation uncertainty, adverse selection, retention, post-acquisition moral hazard, and financing concerns \citep{kohers2000earnouts,datar2001earnouts,cain2011earnouts,cadman2014economic,bates2018financing, barbopoulos2021real}. 
    While these studies can explain why compensation should be contingent, they leave open how the shape of management's contingent compensation affects subsequent performance.
    More broadly, work on executive pay shows that option-like compensation is consistent with increased innovative activity \citep{mao2018managerial}, but does not distinguish effort, risk-taking, or project selection. 

    This distinction suggests testable implications of our model and directions for future empirical study. 
    For one, our model predicts natural comparative statics. Our analysis suggests that greater ex-ante risk in the target's known prospects---in the sense of mean-preserving spreads---should lead to a higher performance threshold, all else equal. Normalizing the threshold by the overall acquisition size would control for potentially different means, while looking across sectors where projects have naturally different volatility could provide the necessary variation.

    \paragraph{Advance-against-royalty contracts.}    
    A similar structure appears in advance-against-royalty contracts, which are common in intellectual-property-generating industries.
    We discuss these contracts in the context of book publishing.
    Under an advance-against-royalty contract, the writer receives an upfront payment (the advance) but does not earn additional royalties until the work generates enough revenue to ``earn out'' the advance \citep{authorsguild2020model}. 
    Above this threshold, the writer becomes a partial residual claimant.
    Thus, after the initial transfer, the writer is effectively placed in a debt-like position until the advance is repaid.
    
    Mapping this scenario into our model, the author would be the ``agent,'' and his initial pitch or provided sample chapters form the basis of the ``known project'' $a_0$.
    The author's creative process can be represented as a directed search,  iteratively investigating alternative storylines or reworking narrative structure after assessing each draft.
    In the end, a single manuscript is delivered to the publisher.

    Our model provides one explanation for the debt-like structure of this contract. By requiring the advance to be worked off, the contract makes modest outcomes relatively unattractive to the author, while royalties above the threshold preserve his incentive to explore risky ideas. That said, our baseline model does not uniquely pin down the structure of the entire contract, such as why the author is only a partial residual claimant of the generated revenue. Two related frictions excluded from our model help explain this feature. First, by retaining a share of the revenue, the publisher is incentivized to take productive actions after the manuscript is delivered, such as advertising, distribution, and negotiating with downstream sellers.\footnote{\citet{poblete2017managing} show that principal moral hazard pushes towards contracts where the principal retains a share of the upside.} Second, royalties are generally based on generated revenue rather than reported profit, making them resistant to unfavorable cost allocation by the publisher. Although the author receives less than the full revenue generated, he may hold a larger claim on the project's actual profits.
    The different components of the contract can therefore serve distinct purposes: the advance provides liquidity, the threshold preserves exploration incentives, and continued revenue sharing provides appropriate publisher incentives.

    There is little work on publishing agreements within the economics literature, suggesting a natural opportunity to study the role of experimentation in a novel setting. 
    The closest evidence comes from a recent study on digital publishing, which uses a different payment model because the writer's progress is tracked incrementally. \citet{LiShiZhao2025MassCreativity} show that making the payment schedule more convex increases the quality of the produced output, directionally aligned with our insights. However, given the different settings and contracting ability, we are hesitant to call this result evidence supporting our main channel. 
    
     The traditional publishing setting offers several empirical advantages relative to others that have been explored. For one, contract terms are generally more transparent and easier to measure than in earnout acquisitions. Projects are also naturally discrete, as are search and stopping decisions. Lastly, there are useful proxies to measure the amount of publisher uncertainty and project risk, such as the author's prior successes and publication genre. Our model offers concrete predictions for how advance size and royalty rate should vary with these factors, providing a basis for future empirical work.

%----------------------------------------%
%----------------------------------------%
%\newpage

\appendix

\section{Proofs Omitted from Main Text}
       
    \subsection{Proof of \Cref{thm:optimaldebtadjacent}}
    \label{proof:thm:optimaldebtadjacent}
    
    \noindent \underline{\textbf{Part 1: MDL \& FSE $\Rightarrow$ $w$ is optimal.}}\quad  We show that $V_P(w) = s_0$ if the contract $w$ satisfies MDL and FSE. That is, the contract $w$ attains the robustly optimal guarantee. Fix the agent's optimal search strategy and the realized prizes $\{y_i\}_{i=0}^n$. Then we consider two cases of the realized search process: (i) $a_0$ is not sampled, and (ii) $a_0$ is sampled.  
    
    Suppose $a_0$ is \textit{not sampled}. The FSE condition guarantees that the $w$-induced index of $a_0$ is exactly 0. Thus, if the agent did not sample $a_0$, the agent must have obtained a strictly positive wage or a wage of zero from a prize that exceeds $s_0$. By the MDL condition, this implies that the agent sampled a project whose prize is higher than $s_0$ and stopped, and, hence, that the principal's payoff is at least $s_0$ after any history such that $a_0$ is \textit{not sampled}. 
    
    Suppose $a_0$ is \textit{sampled}. By the FSE condition, the $w$-induced index of $a_0$ is exactly 0. Because $a_0$ was sampled, the agent could not have drawn a prize with strictly positive wage before sampling $a_0$. Also, after sampling $a_0$, it is optimal for the agent to stop without exploring any additional projects. By favorable tie-breaking, the agent returns either the prize drawn from $a_0$ with a positive wage or the maximum of his prizes with zero wage. In either case, this is superior to sampling $a_0$ alone. Therefore, the principal's expected profit across histories where $a_0$ is sampled is at least $s_0$. 
    
    \noindent \underline{\textbf{Part 2: $w$ is optimal $\Rightarrow$ MDL \& FSE.}}\quad  We prove the contrapositive. First, suppose a contract $w$ violates FSE. Then
    \begin{equation}
    \label{eq:inequality}
    s_0 = V_P(w_0|\mathcal{A}_0) > V_P(w|\mathcal{A}_0) \geq V_P(w), \nonumber
    \end{equation}
    where the equality is proved in Proposition~\ref{thm:debtcontracts}, the first inequality holds strictly because FSE is violated, and the second inequality holds by the definition of $V_P(w)$. Hence, no contract $w$ that violates FSE is robustly optimal as $V_P(w) < s_0$.
    
    Second, suppose a contract $w$ satisfies FSE but violates MDL, i.e., there exists $y'$ such that $y' - w(y') < s_0$ and $w(y')>0$. Consider the set of projects $\mathcal{A} = \{a_0, a_1\}$, where $a_1=(\delta_{y'}, 0)$ is a riskless project. The agent searches only $a_1$ because the $w$-induced index value of $a_0$ is zero by FSE. The principal's corresponding payoff, $y'-w(y')$, is strictly smaller than $s_0$. In other words, $s_0 >  V_P(w|\{a_0, a_1\}) \geq V_P(w)$, and no contract that satisfies FSE but violates MDL is robustly optimal.
    
    Thus, no contract that violates MDL or FSE is optimal, and hence, any optimal contract satisfies MDL and FSE.
    
    \subsection{Proof of \Cref{prop:twobox}}
    \label{pf:prop:twobox}
    
    Suppose $w$ is a robustly optimal contract. Without loss, we let $r^w_0 \geq r^w_1 \geq ... \geq r^w_{k-1}$. 
    
    \noindent \underline{\textbf{Proof of (a).}} \ We prove by induction. Suppose $k=2$, i.e., the principal only knows two projects available to the agent. We begin with the following lemma. 
    
    \begin{lemma}
    \label{lemma:secondbox}
    If $k=2$, at least one of $r^w_0$ and $r^w_1$ must be zero. 
    \end{lemma}
    \begin{proof}
    We prove this lemma by exhausting all possible cases. First, suppose $r^w_1 \leq r^w_0 <0$. In this case, the principal's payoff guarantee under $w$ will be non-positive, as the agent is not willing to explore either known project. 
    
    Second, suppose $r^w_1 < 0 < r^w_0$. In this case, the project $a_1$ has no value to the principal, as the agent never explores it. Therefore, we are back to the case with only one known project, where full surplus extraction of $a_0$ is required for a contract to be robustly optimal, which disallows $r_0^w > 0$. 
    
    Third, suppose $0 < r^w_1 \leq r^w_0$. In this case, construct the following alternative contract $w'(y) = [w(y) - r^w_1]^+$. We show that $w'$ outperforms $w$ in terms of the principal's payoff guarantee. By construction, for any project $a_i$ such that $r^w_i \geq r^w_1$ or $c_i = 0$ (not necessarily in $\mathcal{A}_0$), we have $r_i^{w'} = [r_i^w-r_1^w]^+$. For all other projects $a_i$, we have $r^{w'}_i <0$. Let $\mathcal{A} \supset \mathcal{A}_0$ be any ``adversarial scenario'' against $w'$. We construct another set of projects $\mathcal{A}'$ by removing all the projects in $\mathcal{A}$ that have a $w$-induced index strictly below $r^w_1$ and positive costs. Under such construction, every agent-optimal strategy $\sigma(w, \mathcal{A}')$ given $w$ and $\mathcal{A}'$ is also optimal given $w'$ and $\mathcal{A}$ (since any project removed from $\mathcal{A}$ was never searched under $w'$). So, by the principal-preferred tie-breaking assumption, the principal prefers the agent's strategy given $w'$ and $\mathcal{A}$ to $\sigma(w, \mathcal{A}')$. That is,
    \begin{align*}
        V_P(w' \mid \mathcal{A}) & \geq \mathbb{E}_{\sigma(w, \mathcal{A}')}\left[y - w'(y)\right] \\
        & = \mathbb{E}_{\sigma(w, \mathcal{A}')}\left[y - w(y)\right] + \mathbb{E}_{\sigma(w, \mathcal{A}')}\left[w(y) - w'(y)\right] \\
        & = \mathbb{E}_{\sigma(w, \mathcal{A}')}\left[y - w(y)\right] + \mathbb{E}_{\sigma(w, \mathcal{A}')}\left[ \min \left\{ w(y), r^w_1 \right\} \right].
    \end{align*}
    Observe that (i) if the agent optimally inspects $a_0$, then the agent's wage for the returned prize exceeds the wage he would have obtained from returning the prize generated by $a_0$; and (ii) if, instead, the agent does not search $a_0$, then the agent must have obtained a prize such that his wage exceeds $r^w_0\geq r^w_1$. Thus, 
    \begin{equation*}
        \mathbb{E}_{\sigma(w, \mathcal{A}')}\left[ \min\left\{ w(y), r^w_1\right\} \right] \geq \mathbb{E}_{F_0}\left[\min\left\{w(y), r^w_1\right\}\right] :=\delta >0,
    \end{equation*}
    where the last inequality follows since $r_0^w \geq r^w_1>0$, and, hence, $\mathbb{P} \left( w(y) >0 \right)>0$ and $0< \delta \leq r^w_1$. Therefore, 
    \begin{align*}
        V_P(w' \mid \mathcal{A}) & \geq \mathbb{E}_{\sigma(w, \mathcal{A}')}\left[y - w(y)\right] + \delta \\
        & = V_P(w \mid \mathcal{A}') + \delta.
    \end{align*}
    Since $\delta$ is independent of $\mathcal{A}'$, $V_P(w') > V_P(w)$, which contradicts the optimality of $w$.
    \end{proof}

    Building on \Cref{lemma:secondbox}, we prove \Cref{prop:twobox}(a) for $k=2$ by considering two cases. 

    \paragraph{Case 1: $r_0^w=0$.} \ In this case, the agent must receive zero wage for any $y< V_P$. By contradiction, if there exists $y' < V_P$ such that $w(y') > 0$, then an adversarial project $(\delta_{y'}, 0)$ can crowd out both known projects because it brings a positive wage to the agent, whose $w$-induced indices for both known projects are non-positive. Such an adversarial project yields the principal a payoff strictly lower than $V_P$, a contradiction. 
    
    \paragraph{Case 2: $r_0^w > 0$.} \ In this case, we must have $r_1^w = 0$ according to \Cref{lemma:secondbox}. We let $s_0$ denote the social surplus of project $a_0$, and then we have 
    \begin{equation}
    V_P \geq s_0. \label{eq:twobox_1}
    \end{equation}
    This is because the principal can always secure a payoff guarantee of at least $s_0$ by offering the $r_0$-debt contract. Intuitively, such a contract ensures that the agent is willing to explore $a_0$, and, at the same time, it does not distort the agent's exploration order among projects whose indices are weakly greater than $r_0$. 
    
    Next we show that there exists $\underline{y}$ satisfying Proposition~\ref{prop:twobox}(a).  Suppose, by contradiction, that there does not exist $\underline{y}$ satisfying Proposition~\ref{prop:twobox}(a). Then, for any $r^w_0> \epsilon > 0$, there must exist $y_{\epsilon} < \epsilon$ such that $w(y_{\epsilon}) > 0$. Given $\epsilon$, we consider an adversarial project $a_2 = (\delta_{y_\epsilon}, 0)$. This project will crowd out the project $a_1$ because $r^w_1 = 0$. 
    Hence, the agent's optimal search strategy under $\mathcal{A}=\{a_0, a_1, a_2\}$ is as follows. If $w(y_{\epsilon}) > r^w_0$, he only explores the adversarial project $a_2 = (\delta_{y_\epsilon}, 0)$, which yields a strictly negative payoff to the principal. Otherwise he first explores $a_0$; if $w(y_0) \geq w(y_{\epsilon})$, he will stop exploration and present $y_0$; if $w(y_0) < w(y_{\epsilon})$, he will explore $a_2$ and then stop, completely ignoring $a_1$. In that case, the principal's payoff under $\mathcal{A}=\{a_0, a_1, a_2\}$ is
    \begin{align}
    V_P(w\mid \{a_0, a_1, a_2\}) & = \int (y - w(y)) dF_0(y) + \int_{\{y : w(y) < w(y_{\epsilon})\}} \left[ (y_{\epsilon} - w(y_{\epsilon})) - (y - w(y)) \right] dF_0(y) \notag \\
    & \quad + \int_{\{ y:  w(y) = w(y_{\epsilon}) \}} \left( y_{\epsilon} - y \right)^+ dF_0(y) \label{eq:twobox_2}
    \end{align}
    The first term of (\ref{eq:twobox_2}) is strictly smaller than $s_0$ because $r_0^w>0$ in this case. Let $\Delta:= s_0 - \int (y - w(y)) dF_0(y) > 0$. Meanwhile, the sum of the second and third terms of (\ref{eq:twobox_2}) is smaller than $\Delta$ when $\epsilon$ is sufficiently small. This is because the two integration sets are disjoint, the second integrand satisfies 
    \[(y_{\epsilon} - w(y_{\epsilon})) - (y - w(y)) = (y_{\epsilon} - y) + (w(y) - w(y_{\epsilon})) < y_{\epsilon} - y < \epsilon, \]
    where the first inequality holds because the integration only applies to $\{y : w(y) < w(y_{\epsilon})\}$, and the second inequality is based on $y \geq 0$ and $y_{\epsilon} < \epsilon$, and the third integrand satisfies $(y_{\epsilon} -y)^+ \leq y_{\epsilon} < \epsilon$. Therefore, for sufficiently small $\epsilon$,
    \begin{equation}
        V_P(w\mid \{a_0, a_1, a_2\}) < s_0. \label{eq:twobox_3}
    \end{equation}
    Combining (\ref{eq:twobox_1}) and (\ref{eq:twobox_3}) yields $V_P(w\mid \{a_0, a_1, a_2\}) < V_P$, contradicting the optimality of $w$. 
    
    So far, we have proved Proposition~\ref{prop:twobox}(a) for $k=2$. Now we complete the proof by induction. Suppose Proposition~\ref{prop:twobox}(a) holds for $k = j \geq 2$. We want to show that it also holds for $k=j+1$. 
    
    Suppose $w$ is a robustly optimal contract for $k=j+1$ and continue to assume that $r^w_0 \geq r^w_1 \geq \cdots \geq r^w_j$. Following the same reasoning as in \Cref{lemma:secondbox}, at least one of these indices must be zero, and hence $r^w_j \leq 0$. Suppose, toward a contradiction, that $w$ has no zero-wage region. We show that if there is no zero-wage region, the principal's guarantee under $w$ is unchanged if she knows only the remaining $j$ projects.

    There are two cases. If $r^w_j<0$, project $a_j$ is never explored, so adding it to any set of available projects does not affect the principal's payoff under $w$. Now suppose $r^w_j=0$. Fix any finite set $B$ containing the remaining $j$ known projects. Choose an arbitrarily small $y>0$ with $w(y)>0$, and choose $\varepsilon>0$ arbitrarily small and below both $w(y)$ and every positive $w$-induced index in $B$. Consider the deterministic project $a_\varepsilon=(\delta_y,w(y)-\varepsilon)$.
    This project has induced index $\varepsilon$, so it crowds out $a_j$. Also, because $\varepsilon$ is below every positive $w$-induced index, the same search decisions are optimal in $B$ until $a_\varepsilon$ is explored. If it is explored, it terminates search and gives the principal at most $y$. At the corresponding history in $B$, all remaining projects have non-positive induced indices, so stopping is optimal for the agent. Since the current best prize is nonnegative and its wage is at most $\varepsilon$, stopping gives the principal at least $-\varepsilon$. Favorable tie-breaking then implies
    $$
    V_P(w\mid B\cup\{a_j,a_\varepsilon\})
    \leq V_P(w\mid B)+y+\varepsilon.
    $$
    Since $B$ is arbitrary and $y,\varepsilon$ can be made arbitrarily small, knowing $a_j$ cannot improve the principal's guarantee under $w$. However, knowing $a_j$ also cannot reduce her guarantee because that restricts the possible sets of projects $\mathcal{A}$ available to the agent.
    
    Thus, in either case, the guarantee under $w$ is unchanged when the principal knows only the remaining $j$ projects. Since knowing fewer projects cannot increase her optimal guarantee, $w$ must also be robustly optimal with those $j$ known projects. But, by the induction hypothesis, $w$ has a positive zero-wage region, a contradiction. This completes the induction.
    
    %Now consider the case where $r^w_{j} = 0$, where $a_{j}$ may be explored on some equilibrium path (i.e., the principal finds the project $a_{j}$ potentially valuable). In this case, if we do not have a zero-wage region for a sufficiently small prize, then following the same reasoning as Case 2 above, we can always construct an adversarial project that (i) crowds out $a_{j}$ and (ii) yields an arbitrarily small payoff to the principal. As a consequence of this adversarial scenario, the principal can never generate any value from the project $a_{j}$ in the worst case. This again reduces the principal's profit to the level if she only had $j$ projects available.

    \smallskip
    
    \noindent \underline{\textbf{Proof of (b).}} \ Notice that $w(y) \leq y-V_P$ if $w(y) > r_0^w$; otherwise, an adversarial project $(\delta_{y}, 0)$ will crowd out $a_0$, yielding the principal a payoff strictly less than $V_P$. 
    
    We let $\overline{y}:= r_0^w + V_P$. By contradiction, suppose there exists $y'\geq \overline{y}$ such that $w(y') > y'-V_P$; then it follows that $w(y') > \overline{y} - V_P = r_0^w$, and by the argument in the last paragraph, it follows that $w(y') \leq y'-V_P$, a contradiction.

    \subsection{Proof of \cref{prop:resampling}}
    \label{proof:prop:resampling}

     The optimality of pure debt follows from the same argument as Proposition \ref{thm:debtcontracts}: The pure debt contract $w(y) = \left[y-r^*\right]^+$ (with debt level set equal to the maximal index of all known projects, $r^*$) induces the agent to search until he generates a prize above the debt level $r^*$ for any set of projects $\mathcal{A}\supset \mathcal{A}_0$. So, it guarantees the principal a payoff of at least $r^*$. Since $r^*$ is the maximal total surplus when $\mathcal{A} = \mathcal{A}_0$, $V_P \leq r^*$. Hence, the pure debt contract $w(y) = \left[y-r^*\right]^+$ is robustly optimal. 

    To establish $F^*$-uniqueness, observe first that for any other contract $w'$ to achieve the payoff guarantee of $r^*$, we must have $w'(y) \leq [y-r^*]^+$ for all $y$. If not, there exists some $y$ such that $w'(y) > [y-r^*]^+ \geq y - r^*$, so $y - w'(y) < r^*$. This means adding a free, constant-output project with prize $y$ would reduce the principal's payoff below $r^*$, and hence $w'$ would have a lower payoff guarantee. Further, for any contract $w'$ to achieve the same payoff guarantee as $w$, it also needs to extract the agent's full surplus, meaning $\mathbb{E}_{F^*}[w'(y)] = c^*$. But, by construction of the index, $\mathbb{E}_{F^*}[w(y)] = \mathbb{E}_{F^*}[(y-r^*)^+] = c^*$. Since $w'$ is pointwise below $w$ and it has the same expectation, they can only differ on an $F^*$-null set.
    
    \subsection{Proof of \Cref{prop:tranche}}
    \label{proof:prop:tranche}

    The proof is divided into three steps. First, we bound the principal's guarantee from above. Second, we show that a capped-earnout debt contract $w^*(y)$ achieves this upper bound. Finally, we prove that any robustly optimal contract coincides with $w^*(y)$ $F_0$-almost everywhere.

    \noindent \textbf{Step 1:} For all $w$, $V_P(w) \leq V$, where $0 \leq V \leq \bar{V}$ is the unique fixed point of the mapping
    \begin{align*}
        G: v \in [0, \bar{V}] \to G(v) = \max_{\tilde{w}}\left\{ \mathbb{E}_{F_0}\left[ y -\tilde w(y) \right] \ : \ \mathbb{E}_{F_0}[u(\tilde w(y))] = c_0 \ \& \ 0 \leq \tilde w(y) \leq (y-v)^+\right\}
    \end{align*}
    given by Lemma \ref{lemma:FP_riskaversion} and $\bar{V}$ is the smallest solution of $\mathbb{E}_{F_0}\left[u\left([y-\bar{V}]^+\right)\right] = c_0$. Note that $\bar{V}<\infty$ since $\mathbb{E}_{F_0}\left[u\left(y\right)\right] > c_0>0.$

    Let $w$ be a contract. If $\mathbb{E}_{F_0}[u(w(y))] < c_0$, $V_P(w) \leq 0 \leq V$ and we are done. So, assume that $\mathbb{E}_{F_0}[u(w(y))] \geq c_0$. Since the adversarial nature can choose a set of projects consisting of the known project $(F_0, c_0)$ and a cheap and safe alternative $(\delta_{\{y\}}, 0)$ with $y\geq 0$, 
    \begin{align*}
        V_P(w) \leq \min\left\{ S(w), \ \mathbb{E}_{F_0}\left[ y- w(y)\right] \right\}.
    \end{align*}
    Here, $S(w)$ is the principal's payoff when the worst possible ``cheap and safe'' alternative crowds out the known project:
    \begin{align*}
        S(w) = \inf_{y\geq 0}\left\{ y - w(y) \ : \ u(w(y)) > r_0^w \right\},
    \end{align*}
    where $r_0^w$ is the wage-induced index of the known project. We show that
    \begin{align*}
        \min\left\{ S(w), \ \mathbb{E}_{F_0}\left[ y- w(y)\right] \right\} \leq V.
    \end{align*}
    Observe first that there exists a contract $w'$ with $\mathbb{E}_{F_0}[u(w'(y))] = c_0$ such that
    \begin{align*}
       \min\left\{ S(w), \ \mathbb{E}_{F_0}\left[ y- w(y)\right] \right\} \leq \min\left\{ S(w'), \ \mathbb{E}_{F_0}\left[ y- w'(y)\right] \right\}.
    \end{align*}
    This follows because when $\mathbb{E}_{F_0}[u(w(y))] > c_0 \Leftrightarrow r_0^w >0$, the contract $w'$ defined by 
    \begin{align*}
        u(w'(y)) = \left(u(w(y)) -r^w_0\right)^+
    \end{align*}
    induces a zero $w'$-induced index for the known box, $r^{w'}_0 =0$. Moreover, $w'$ is pointwise smaller than $w$ since $u$ is strictly increasing and, by definition, $\left\{ y \ : \  u(w(y)) > r_0^w \right\} = \left\{ y \ : \  u(w'(y)) > r_0^{w'} \right\}$. So, $S(w) \leq S(w')$ and $\mathbb{E}_{F_0}\left[ y- w(y)\right] \leq \mathbb{E}_{F_0}\left[ y- w'(y)\right]$. Therefore,  
    \begin{align*}
       \min\left\{ S(w), \ \mathbb{E}_{F_0}\left[ y- w(y)\right] \right\} \leq \min\left\{ S(w'), \ \mathbb{E}_{F_0}\left[ y- w'(y)\right] \right\}.
    \end{align*}
    It remains to show that $\min\left\{ S(w'), \ \mathbb{E}_{F_0}\left[ y- w'(y)\right] \right\} \leq V$. If $S(w') \leq V$, we are done. So, suppose that $S(w') > V$. The definition of $S$ then implies that $y-w'(y) > V$ whenever $u(w'(y)) >0 \Leftrightarrow w'(y) >0$. So, $0 \leq w'(y) \leq (y-V)^+$ for all $y$. As a result, the contract $w'$ is feasible in the maximization problem $G(V)$ and
    \begin{align*}
        \mathbb{E}_{F_0}\left[ y- w'(y)\right] \leq \max_{\tilde{w}}\left\{ \mathbb{E}_{F_0}\left[ y -\tilde w(y) \right] \ : \ \mathbb{E}_{F_0}[u(\tilde w(y))] = c_0 \ \& \ 0 \leq \tilde w(y) \leq (y-V)^+\right\} = G(V).
    \end{align*}
    But, $V$ is a fixed point of $G(V)$. So, $V_P(w) \leq V$.

    \noindent \textbf{Step 2:} The capped-earnout debt contract maximizing $G(V)$ given by Lemma \ref{lemma:FP_riskaversion}, $w^*(y) = \min\left\{ [y-V]^+, C_V\right\}$ (where $C_V\geq 0$ solves $\mathbb{E}_{F_0}\left[ u\left(\min\left\{ [y-V]^+, C_V\right\}\right)\right] =c_0$), is robustly optimal. 

    By Step 1, $V_P(w^*) \leq V_P \leq V $. We show that $V_P(w^*) \geq V$. Let $\mathcal{A}$ be an arbitrary set of projects that contains $(F_0, c_0)$. Given contract $w^*$, if the agent samples $(F_0, c_0)$, the principal's payoff is at least
   \begin{align*}
       \mathbb{E}_{F_0}\left[ y- w^*(y) \right] = G(V) =V,
   \end{align*}
   where we used that $V$ is the unique fixed point of $G$. If, instead, the agent does not sample $(F_0, c_0)$, then he must have terminated the search with a positive wage or returned a prize of at least $V$ to the principal when getting a zero wage (for otherwise he would optimally sample $(F_0, c_0)$ under principal-preferred tie-breaking). Therefore, for the returned prize $\tilde{y}$,  either $w^*(\tilde{y})>0$ which implies $\tilde{y} - w^*(\tilde{y}) \geq V$, or $\tilde{y} \geq V$ and $w^*(\tilde{y}) =0$. In both cases, the principal's payoff is at least $V$. 
   
   Since $\mathcal{A}$ was arbitrary, we have $V_P(w^*\mid \mathcal{A}) \geq V$ for all $\mathcal{A} \supset \{(F_0, c_0)\}$. Hence, $V_P(w^*) \geq V$, and $w^*(y) = \min\left\{ [y-V]^+, C_V\right\}$ is robustly optimal and $V_P = V_P(w^*) = V$.
   
   \noindent \textbf{Step 3:} Any robustly optimal contract coincides with $w^*(y)$ $F_0$-almost everywhere.

   Let $w$ be a robustly optimal contract. Then $V_P(w) = V >0$ and, hence, $\mathbb{E}_{F_0}[u(w(y))] \geq c_0$. Since nature can, as in Step 1, choose a set of projects that consists of the known project and a cheap and safe alternative, we have
   \begin{align*}
       V_P(w) \leq \min\left\{ S(w), \ \mathbb{E}_{F_0}\left[ y- w(y)\right] \right\}.
   \end{align*}
   Following the argument in Step 1, if $\mathbb{E}_{F_0}[u(w(y))] > c_0$, there exists a pointwise smaller contract $w'$ with $\mathbb{E}_{F_0}[u(w'(y))] = c_0$ such that
   \begin{align*}
       V_P(w) =V  \leq \min\left\{ S(w), \ \mathbb{E}_{F_0}\left[ y- w(y)\right] \right\} \leq \min\left\{ S(w'), \ \mathbb{E}_{F_0}\left[ y- w'(y)\right] \right\}.
   \end{align*}

   If $\mathbb{E}_{F_0}[u(w(y))] = c_0$, simply set $w'=w$.
   But $S(w') \geq V$ implies that $0\leq w'(y) \leq [y-V]^+$ for all $y$. So, $w'(y)$ is feasible in the maximization problem $G(V)$, and, hence, optimal, as nature can give the agent only the known project:
   \begin{align*}
       \mathbb{E}_{F_0}\left[ y- w'(y)\right] \geq V_P(w) =  V = G(V).
   \end{align*}
   Lemma \ref{lemma:FP_riskaversion} then guarantees that $w'(y) = w^*(y)$ for $F_0$-almost every $y$. Since $w'(y) \leq w(y)$, this in turn implies that
   \begin{align*}
       V \leq \mathbb{E}_{F_0}\left[ y- w(y)\right] \leq  \mathbb{E}_{F_0}\left[ y- w'(y)\right] = V.
   \end{align*}
   Therefore, $w(y) = w'(y) = w^*(y)$ $F_0$-almost everywhere. \hfill \qed 
   
    \begin{lemma}\label{lemma:FP_riskaversion}
        Let $\bar{V}$ be the smallest solution of $\mathbb{E}_{F_0}\left[u\left([y-\bar{V}]^+\right)\right] = c_0$. For all $v \in [0, \bar{V}]$, the capped-earnout debt contract $w^*_v(y) = \min\left\{ [y-v]^+, C_v \right\}$, with $C_v$ defined by\footnote{When $v= \bar{V}$, $C_v$ may be $\infty$.}
        \begin{align*}
            \mathbb{E}_{F_0}\left[u\left(\min\left\{ [y-v]^+, C_v \right\}\right)\right] = c_0
        \end{align*}
        is the $F_0$-unique solution of the maximization problem
        \begin{align*}
            G(v) = \max_{\tilde{w}}\left\{ \mathbb{E}_{F_0}\left[ y -\tilde w(y) \right] \ : \ \mathbb{E}_{F_0}[u(\tilde w(y))] = c_0 \ \& \ 0 \leq \tilde w(y) \leq [y-v]^+\right\}.
        \end{align*}
        Moreover, $G(v)$ is non-increasing, continuous, and has a unique fixed point $V \in  [0, \bar{V}]$.
    \end{lemma}

    \begin{proof}
        We first prove that the capped-earnout debt contract $w^*_v(y) = \min\left\{ [y-v]^+, C_v \right\}$ $F_0$-uniquely solves $G(v)$ for all $v \in [0, \bar{V}]$.
    
        Let $v \in [0, \bar{V}]$ and consider the maximization problem $G(v)$. If $C_v = \infty$ (which implies $v= \bar{V}$), the strictly increasing function $w \mapsto u(w)$ is uniquely maximized by $ w(y) = [y-v]^+$ on $[0, [y-v]^+]$ for all $y$, and, hence $w(y) = [y-\bar{V}]^+$ $F_0$-uniquely solves $G(\bar{V})$. If $C_v < \infty$, the strict concavity of $u$ implies that\footnote{If $u$ is not differentiable at $C_v$, we take $u'(C_v)$ in the supergradient set of $u$.} $w \mapsto u(w) - u'(C_v) w$ is uniquely maximized by $ w(y) = \min\left\{ [y-v]^+, C_v\right\}$ on $[0, [y-v]^+]$ for all $y$. Therefore, for all $y$ and all $0\leq w(y) \leq (y-v)^+$, 
        \begin{align*}
            u\left(\min\left\{ (y-v)^+, C_v\right\}\right) - u'(C_v) \min\left\{ (y-v)^+, C_v\right\} \geq  u(w(y)) - u'(C_v) w(y),
        \end{align*}
        and the inequality is strict when $w(y) \neq \min\left\{ (y-v)^+, C_v\right\}$. Taking expectations with respect to $F_0$, we obtain
        \begin{align*}
            \mathbb{E}_{F_0}\left[ u\left(\min\left\{ [y-v]^+, C_v\right\}\right) \right] - \mathbb{E}_{F_0}\left[  u(w(y)) \right] \geq u'(C_v) \left( \mathbb{E}_{F_0}\left[ \min\left\{ [y-v]^+, C_v\right\} \right] - \mathbb{E}_{F_0}\left[  w(y) \right] \right).
        \end{align*}
        Since $\mathbb{E}_{F_0}\left[ u\left(\min\left\{ [y-v]^+, C_v\right\}\right) \right] = \mathbb{E}_{F_0}\left[  u(w(y)) \right] = c_0$ and $u'(C_v) >0$, \ $\mathbb{E}_{F_0}\left[ \min\left\{ [y-v]^+, C_v\right\} \right] \leq  \mathbb{E}_{F_0}\left[  w(y) \right]$, with strict inequality if $w(y) \neq  \min\left\{ [y-v]^+, C_v\right\}$ on an $F_0$-positive probability set. Therefore, $w^*_v(y) = \min\left\{ [y-v]^+, C_v\right\}$ is the $F_0$-unique minimizer of $ w(y) \mapsto \mathbb{E}_{F_0}[w(y)]$ subject to $\mathbb{E}_{F_0}\left[ u(w(y)) \right] = c_0$ and $0\leq w(y) \leq [y-v]^+$. Hence, it is the $F_0$-unique solution of the  maximization problem $G(v)$ and 
        \begin{align*}
            G(v) = \mathbb{E}_{F_0}\left[y- \min\left\{ [y-v]^+, C_v \right\}\right].
        \end{align*}

        It remains to show that $G(v)$ is continuous, nonincreasing, and has a unique fixed point. Monotonicity follows from the definition of $G(v)$. For $\underline{v} < \bar{v}$, $w^*_{\bar{v}}$ is feasible in the maximization problem $G(\underline{v})$ whose unique maximizer is $w^*_{\underline{v}}$. Hence, $G(\underline{v}) \geq G(\bar{v})$. Continuity follows from the continuity of $v\mapsto C_v$ and the dominated convergence theorem. Therefore $G(v)$ has a unique fixed point on $[0, \bar{V}]$ by the intermediate value theorem since $G(\bar{V}) = \mathbb{E}_{F_0}\left[ \min \left\{ y, \bar{V}\right\}\right] \leq \bar{V}$ and $G(0) \geq 0$. 
    \end{proof}

%------------------%
%------------------%

\section{Earnout Contract Example}
\label{appendix:earnout_example}

We provide an example of an earnout agreement that exactly takes the capped-earnout debt contract form characterized in Section~\ref{subsec:riskaverse}, \cref{prop:tranche}. The example comes from Applied Signal Technology, Inc.'s 2009 acquisition of Pyxis Engineering LLC. The agreement is publicly available on the SEC website. The acquisition agreement specifies an earnout based on revenues during the twelve months following the acquisition. In particular, it states:
\begin{quote}
2.6(b). \ No Earn-Out Distribution Amount shall be payable unless the Target Revenues exceed Thirteen Million Two Hundred Fifty Thousand Dollars (\$13,250,000) (the `Threshold Target Revenues'). The Earn-Out Distribution Amount shall equal every dollar of Target Revenues that exceed the Threshold Target Revenues, up to a maximum Earn-Out Distribution Amount of Three Million Seven Hundred Fifty Thousand Dollars (\$3,750,000) (the `Maximum Earn-Out Distribution Amount') if Target Revenues reach Seventeen Million Dollars (\$17,000,000) or more.
\end{quote}
\noindent
Thus, letting $y$ denote Target Revenues (the revenue generated by the acquired firm), the contingent payment to the acquired firm is
\[
    w(y)
    =
    \min\left\{
        \$3.75\text{m},
        [y-\$13.25\text{m}]^+
    \right\}.
\]
This is exactly the capped-earnout debt contract in Section~\ref{subsec:riskaverse}, with debt level $z=\$13.25$ million and wage cap $\bar w=\$3.75$ million.%
\footnote{Applied Signal Technology, Inc., Membership Interest Purchase Agreement with Pyxis Engineering LLC, filed with the U.S.\ Securities and Exchange Commission on September 4, 2009, Section 2.6(b). See \citet{appliedsignal2009pyxis}.}

%------------------%
%------------------%

\bibliography{biblio}

%------------------%
%------------------%

\newpage
\section{Online Appendix: Bayesian Benchmark}
\label{apx:bayesian}

This appendix studies a Bayesian version of our problem motivated by the adversarial scenarios we identify. The benchmark is kept simple to maintain tractability and clarify the economic similarities and differences between our robust approach and the Bayesian formulation. 

\subsection{Setup} 

The agent has two projects. The first project $a_0=(F_0, c_0)$ is known to the principal. The second project $a_1 = (\delta_x, 0)$ is a cheap and safe project. The principal only knows that $x \sim G$ but is uncertain about its realization. For simplicity, we assume that both $F_0$ and $G$ have full support on $[0, \infty)$ and are atomless, and $c_0>0$. Moreover, we restrict attention to doubly monotone contracts, so that both $w(y)$ and $y-w(y)$ are non-decreasing.

\subsection{Key Tradeoff: Surplus Generation vs. Rent Minimization}

Since $a_1$ is free, the agent always explores it first. If $w(x) > r_0^w$, the agent terminates his search after exploring $a_1$; if $w(x) \leq r_0^w$, he explores both projects and returns $\max\{y_0, x\}$. Let $z(w):=\sup\{x\mid w(x) \leq r_0^w\}$ be the threshold value of $x$ that induces the agent to search both projects. 

\begin{lemma}
\label{lemma:bayesian}
Either the zero contract ($w \equiv 0$) is optimal, or every optimal contract $w$ satisfies $r_0^w = 0$. 
\end{lemma}
\begin{proof}
Suppose the zero contract is not optimal. Let $w$ be an optimal contract. We show that $r^w_0 =0$ by contradiction. If $r_0^w < 0$, then the agent only explores $a_1$, which indicates that the zero contract must weakly dominate $w$, a contradiction. If $r_0^w > 0$, then we construct $w' = [w-r_0^w]^+$. It does not change the exploration decision of the agent (because $z(w) = z(w')$) and leaves the agent with a strictly lower wage. So $w$ is not optimal, another contradiction. Thus, either the zero contract is optimal, or any optimal contract $w$ satisfies $r_0^w = 0$. 
\end{proof}

We refer to a contract that satisfies $r_0^w=0$ as an $a_0$-FSE contract. \Cref{lemma:bayesian} suggests a useful way to solve this benchmark. We first solve the problem within the class of $a_0$-FSE contracts, and then compare the optimal $a_0$-FSE contract with the zero contract. 

\paragraph{\underline{Optimal $a_0$-FSE contract.}} Facing an $a_0$-FSE contract $w$, the agent explores both projects if and only if $x\leq z(w)=\sup\{x\mid w(x) = 0\}$. When $x\leq z(w)$, the agent's expected payoff is zero, since he earns zero wage from $a_1$ and obtains zero expected payoff from $a_0$; when $x > z(w)$, he receives a wage $w(x)$, which we interpret as his rent from owning a good project $a_1$. Hence, the principal's expected payoff from such a contract $w$ is
\begin{align}
    \Pi(w)
    &=
    \int_0^{z(w)}
    \left[
        \mathbb{E}_{F_0}\bigl(\max\{y_0,x\}\bigr)-c_0
    \right]dG(x)
    +
    \int_{z(w)}^\infty x\,dG(x)
    -
    \int_{z(w)}^\infty w(x)\,dG(x)
    \label{eq:bayesian-payoff}\\
    &=:\pi(z(w))-\int_{z(w)}^\infty w(x)\,dG(x). \nonumber
%    \label{eq:bayesian-decomp}
\end{align}
The first two terms of \Cref{eq:bayesian-payoff}, which we denote as $\pi(z(w))$, are the social surplus generated by the induced search rule. The third term of \Cref{eq:bayesian-payoff} is the agent's rent. The principal's problem can therefore be formulated as 
\begin{eqnarray}
    \sup_{w \text{ is doubly monotone }} && \pi(z(w))-\int_{z(w)}^\infty w(x)\,dG(x) \nonumber \\
    s.t. && \int_{z(w)}^{\infty}w(y) dF_0(y) = c_0. \nonumber
\end{eqnarray}

To highlight the principal's tradeoff in this problem, we solve it in two steps. In the first step, we fix $z(w) = z$ and solve the following rent minimization problem. 
\begin{eqnarray}
    \inf_{w \text{ is doubly monotone }} && \int_{z}^{\infty}w(x)dG(x) \nonumber \\
    s.t. && \int_{z}^{\infty}w(y) dF_0(y) = c_0, z(w)=z. \nonumber
\end{eqnarray}
Let $R(z)$ be the infimum value of this rent-minimization problem. $R(z)$ weakly increases in $z$, because achieving a higher $z$ requires the contract to postpone compensation to higher realizations, making it harder to satisfy the $a_0$-FSE condition. Hence, we can now proceed to the second step, where we solve the following problem. 
\begin{equation}
\sup_{z\in[0, r_0]} \quad \pi(z) - R(z)     \label{eq:bayesian_tradeoff}
\end{equation}
In this problem, $z\leq r_0$ is necessary because under any doubly monotone contract, the agent's marginal share of revenue is weakly lower than one, inducing him to search less than is socially optimal -- i.e., exploring $a_0$ if and only if $x \leq r_0$. 

This problem decomposes the principal's objective into two parts: maximizing social surplus, $\pi(z)$, and minimizing the agent's rent, $R(z)$. We can infer that $\pi(z)$ increases in $z \in[0, r_0]$, as a larger $z$ in this range indicates less distortion from the social optimum. This realization, combined with the aforementioned statement that $R(z)$ weakly increases in $z$, enables us to identify the principal's tradeoff in this problem: a higher $z$ generates more social surplus but also entails a higher rent for the agent. Thus, the solution $z^*$ of problem \eqref{eq:bayesian_tradeoff} must optimally balance the principal's dual objectives of surplus generation and rent minimization. 

\paragraph{\underline{Solution to the Bayesian benchmark.}} We now state the full solution to the Bayesian benchmark. Let $\Pi^* = \pi(z^*) - R(z^*)$ be the principal's optimal payoff from using an $a_0$-FSE contract. Notice that her payoff from using a zero contract ($w\equiv 0$) is $\mathbb{E}_G[x]$. 

\begin{proposition}
\label{prop:bayesian-characterization}
If $\Pi^*>\mathbb{E}_G[x]$, any optimal $a_0$-FSE contract is optimal. If $\Pi^*<\mathbb{E}_G[x]$, the zero contract is optimal. If $\Pi^*=\mathbb{E}_G[x]$, both the zero contract and the optimal $a_0$-FSE contracts are optimal. 
\end{proposition}
\begin{proof}
It immediately follows from \Cref{lemma:bayesian}. 
\end{proof}

\subsection{Features of Optimal Contracts}

Based on \Cref{prop:bayesian-characterization}, we further analyze some features of the optimal contract. 

\begin{proposition}
\label{prop:bayesian-positive-debt}
If the zero contract is not optimal, every optimal contract contains strictly positive debt, i.e., $z^*>0$ always holds.
\end{proposition}
\begin{proof}
By contradiction, suppose $z^*=0$. Let $w$ be an $a_0$-FSE contract satisfying $z(w)=0$. By definition, we know that $\pi(z(w)) = \pi(0) = \mathbb{E}_G[x]$. Meanwhile, we have $R(z(w)) > 0$, because $c_0 > 0$ and the agent must be paid a positive wage for some value of $x$ to be willing to explore $a_0$. Therefore, we infer that $z^*=0$ necessarily implies that $\pi(z(w)) - R(z(w)) < \mathbb{E}_G[x]$, and consequently, any $a_0$-FSE contract must be dominated by the zero contract. 
\end{proof}

\Cref{prop:bayesian-positive-debt} implies the following: within the class of doubly monotone contracts, the optimal contract is either the zero contract or one with strictly positive debt. It complements the insight generated by our model and suggests that a positive debt level is not an artifact of the robust approach. Even when the principal has a Bayesian prior over the unknown safe alternative, she optimally withholds all compensation for sufficiently low realizations. Intuitively, this is because exploring the risky project is most valuable when the agent's safe option is the smallest. Hence, it is most costly---from a total surplus perspective---to induce the agent to settle for a safe, low-value alternative rather than continue searching. This is the same force underlying our robust model.

Next, we examine what happens beyond the debt level of an optimal $a_0$-FSE contract. Unlike in the robust model, the shape of this contract is not distribution-free, as it generally depends on the statistical features of $F_0$ and $G$. To see this, consider the first step of the optimization problem above: for a fixed exploration threshold $z$, the principal chooses an $a_0$-FSE contract that minimizes the agent's information rent. Since $w$ is doubly monotone, it can be written as
\[
    w(y)=\int_0^y q(t)\,dt,
    \qquad
    q(t)\in[0,1].
\]
By Fubini's theorem, the $a_0$-FSE condition becomes
\[
    \int_0^\infty q(t)[1-F_0(t)]\,dt=c_0,
\]
while the agent's expected rent from the safe project is
\[
    \int_0^\infty q(t)[1-G(t)]\,dt.
\]
Hence, the rent-minimization problem is a linear optimization problem. The relevant statistic is the relative-tail ratio
\[
\rho(t):=\frac{1-F_0(t)}{1-G(t)}.
\]
The solution is bang-bang in the sense that there exists a cutoff $\rho^*$ such that\footnote{The cutoff is uniquely pinned down by
$\int_{\{t:\rho(t)>\rho^*\}}[1-F_0(t)]\,dt
\leq c_0 \leq
\int_{\{t:\rho(t)\geq\rho^*\}}[1-F_0(t)]\,dt$.}
\[
q(t)=
\begin{cases}
1, & \rho(t)>\rho^*,\\
0, & \rho(t)<\rho^*.
\end{cases}
\]
Therefore, the optimal contract can arbitrarily alternate between flat regions and slope-1 regions beyond the debt level; importantly, it need not resemble any contracts we observe in practice. Nevertheless, we can impose additional restrictions on $\rho(t)$ to make specific debt-like instruments stand out in the Bayesian model, as suggested by the following proposition. 

\begin{proposition}
\label{prop:bayesian-pure-debt}
Suppose the ratio $\rho(t)$ is weakly increasing in $t$. Then, the pure debt contract $w_0$ is optimal among all $a_0$-FSE contracts. Consequently, the Bayesian optimum is either the zero contract or $w_0$.
\end{proposition}
\begin{proof}
When $\rho(t)$ is weakly increasing, for any given $\rho^*$, $\rho(t) > \rho^*$ if and only if $t$ is above a certain threshold, pointing to the pure debt contract $w_0$. Meanwhile, $w_0$ also maximizes social surplus. Putting the two arguments together, we know that $w_0$ simultaneously generates the highest social surplus and the lowest rent among $a_0$-FSE contracts, and is therefore optimal within that class. 
\end{proof}

The Bayesian benchmark therefore preserves the incentive to withhold payment at low outputs, while adding a distribution-dependent tradeoff between efficiency and information rent. When the relative tail ratio $\rho(t)$ is increasing, these forces align in favor of pure debt.

\end{document}